\documentclass[aps,twocolumn]{revtex4}
\usepackage[dvips]{graphicx}
\def\be{\begin{equation}}
\def\ee{\end{equation}}
\def\bea{\begin{eqnarray}}          
\def\eea{\end{eqnarray}}
\def\bi{\begin{itemize}}
\def\ei{\end{itemize}}

\begin{document}

\title{Solitons in coupled atomic-molecular 
Bose-Einstein condensates}

\author{ Bart\l{}omiej Ole\'s and Krzysztof Sacha }

\address{ Marian Smoluchowski Institute of Physics and 
Mark Kac Complex Systems Research Centre, Uniwersytet Jagiello\'nski,
Reymonta 4, 30-059 Krak\'ow, Poland  }

\date{\today}

\begin{abstract}
We consider coupled atomic-molecular Bose-Einstein condensate system in a 
quasi-one-dimensional trap. In the vicinity of a Feshbach resonance
the system can reveal parametric soliton-like behavior. We analyze bright 
soliton, soliton train and dark soliton solutions for the system in the 
trap and in the presence of the interactions between particles
and find the range of the system parameters where the soliton states can 
be experimentally prepared and detected. 
\end{abstract}

\maketitle

\section{Introduction}

The experimental success in realizing a Bose-Einstein condensate (BEC)
in dilute atomic gases \cite{bec_rb87, bec_na23, bec_li7} has led to
the strong experimental and theoretical activity in this field.
Dilute gases turn out to be an ideal 
system for realizing and high accuracy manipulation of various quantum 
many-body phenomena. The nonlinear nature of a BEC in the mean field 
description allows investigations of e.g. 
vortex lattices \cite{vl_1st_obs,ketterle,vort_latt_str}, 
four-wave mixing \cite{fw_mix_exp, fw_mix_theory},
Josephson-like oscillations \cite{jos_theory, jos_exp}
and solitons in the system.
The latter have been realized experimentally for a quasi one-dimensional (1D) 
atomic BEC with repulsive effective interactions (so-called dark solitons)
\cite{burg, phil} and with attractive interactions (bright solitons) 
\cite{sal,trains_li}. There are also proposals to prepare experimentally 
solitons in a mixture of bosonic and fermionic atoms \cite{trainsBF} 
and in an atomic BEC with a time-dependent scattering length 
\cite{Kevrekidis,abdullaev,pelinovsky,sol_time_scatt,trippenbach}.

In the present publication we consider another type of solitonic solutions. 
That is, parametric solitons, which in the context of nonlinear optics
occur as a result of a coupling between fields in a nonlinear medium
such that each field propagates as solitons \cite{kivshar}. 
In the coupled atomic-molecular BEC, there is a
nonlinear resonant transfer between atoms and molecules,
as well as terms proportional to the densities.
The solitons of this type have been investigated in nonlinear optics 
\cite{karpierz95, drummond96, berge} and in the problem of 
the self-localization of impurity atoms in a BEC \cite{st06}.
In the context of the atomic-molecular BECs the parametric solitons
have been analyzed in Ref.~\cite{drummondPRL98,drummondPRA04}. The latter
publications consider, in principle, solitons in free space in any dimension 
but they concentrate on the existence and stability of the soliton like 
states in 3D free space. However, a profound analysis of the system in a 
quasi 1D trap is missing and methods for experimental preparation 
and detection are not considered. Trapping an atomic BEC in a quasi 1D 
potential allowed realizing
remarkable experiments where bright soliton wave packets travelled in 1D 
space (still with transverse confinement) without spreading \cite{sal,trains_li}. 
In the case of the coupled atomic-molecular BECs, 
experimentalists have to face a problem of atomic losses,
which becomes significant at a Feshbach resonance.
We show that to overcome 
this obstacle one has to deal with moderate particle densities, and consider 
the range of the system parameters where preparation and detection of the 
solitons are sufficiently fast to be able to compete with the 
loss phenomenon. Moreover, the soliton states can be prepared at 
the magnetic field slightly off the resonance value, where the
number of created molecules is still considerable but the rate of atomic 
losses much smaller than at the resonance.

The paper is organized as follows.
In Sec.~II we introduce the model of the system and in Sec.~III we analyze different 
solitonic states. Section~IV is devoted to analysis of methods
for experimental preparation and detection.

\section{The Model}

In order to obtain a complete model of a Feshbach resonance in an atomic 
BEC, the intermediate bound states (molecules) have to be included explicitly 
in the Hamiltonian. At the resonance, the number of molecules becomes considerable
and a second (molecular) BEC is formed. We take into account two body 
atom-atom, molecule-molecule and atom-molecule collisions,
as well as the term responsible for the creation of molecules, i.e. transfer of pairs 
of atoms into molecules and vice versa \cite{timm}:
\bea
	\hat{H} &=& \large{\int} {\rm d}^3 r \Big( \hat{\psi}^{\dagger}_a \Big[ -\frac{\hbar^2}{2m}\nabla^2 
	+ U_a(\vec r) + \frac{\lambda_a}{2} \hat{\psi}^{\dagger}_a \hat{\psi}_a \Big] \hat{\psi}_a \nonumber \\
	&+& \hat{\psi}^{\dagger}_m \Big[ -\frac{\hbar^2}{4m}\nabla^2  
	+ U_m(\vec r) + {\cal E} + \frac{\lambda_m}{2} \hat{\psi}^{\dagger}_m \hat{\psi}_m \Big] \hat{\psi}_m \cr \nonumber \\
	&+& \lambda_{am} \hat{\psi}^{\dagger}_a  \hat{\psi}_a  \hat{\psi}^{\dagger}_m  \hat{\psi}_m
	+ \frac{\alpha}{\sqrt{2}} \big[ \hat{\psi}^{\dagger}_m \hat{\psi}_a \hat{\psi}_a 
	+ \hat{\psi}_m \hat{\psi}^{\dagger}_a \hat{\psi}^{\dagger}_a \big] \Big),
	\label{HFR}	
\eea
where $m$ is the atomic mass, $\lambda_a$, $\lambda_m$, $\lambda_{am}$ 
are coupling constants for 
the respective interactions, and $\alpha$ determines the strength of 
the resonance. The $\cal E$ parameter is a difference between the bound state 
energy of two atoms and the energy of a free atom pair \cite{timm}, and it
can be varied by means of a magnetic field. $U_a(\vec r)$ and 
$U_m(\vec r)$ stand for atomic and molecular trapping potentials, 
respectively. We would like to point out that in our model an internal 
structure of molecules and all processes involving the internal structure are 
neglected and we consider only a single molecular state described by the 
operator $\hat{\psi}_m$. 

In the present paper we consider the system parameters mainly 
related to $^{87}$Rb atoms. That is, the mass $m$, the coupling constant 
$\lambda_a$ and the strength of the resonance $\alpha$ correspond to $^{87}$Rb 
atoms and the Feshbach resonance that occurs at the magnetic field of 
$B_{\rm r} = 685.43$~G \cite{durrprec}. However, because the precise values 
of the coupling constants $\lambda_{m}$ and $\lambda_{am}$ are unknown 
these constants have to be chosen arbitrary. 
For simplicity we assume $\lambda_{am}\approx\lambda_m\approx \lambda_a = 
4 \pi \hbar^2 a / m$, where $a = 5.7$~nm is the value of the atomic scattering 
length far from a Feshbach resonance. In Sec.~III we show that this
assumption is not essential because in a broad range of $\lambda_{m}$ and 
$\lambda_{am}$ values there exist stable soliton solutions.
The resonance strength parameter 
$\alpha$ can be expressed in terms of $\Delta B$,
the resonance width, and $\Delta \tilde{\mu}$, difference between magnetic 
moments of a molecule and a free atom pair, 
$\alpha = \sqrt{ 4 \pi \hbar^2 a \Delta \tilde{\mu} \Delta B/m}$.
We have chosen for investigation the broad resonance which occurs at 
the magnetic field $B_{\rm r} = 685.43$~G where $\Delta B = 0.017 $~G and 
$\Delta \tilde{\mu} = 1.4 \mu_{B}$ ($\mu_B$ is the Bohr magneton) \cite{durrprec}.
We focus on the case where the system is prepared in a harmonic trap with so strong 
radial confinement that the radial frequency of the trap exceeds the chemical potential 
of the system. Then, only the ground states of the transverse degrees of freedom are 
relevant and the system becomes effectively 1D. The chosen trap frequencies
are $\omega_{m,\perp}=\omega_{a,\perp}=2\pi\times 1500$~Hz and
$\omega_{m,x}=\omega_{a,x}=2\pi\times 10$~Hz
\footnote{The same trap frequencies for atoms and molecules are chosen for
simplicity. Different frequencies in the transverse directions would lead only 
to modification of the effective coupling constants in the 1D 
Hamiltonian (\ref{HFR}). Different frequencies in the longitudal direction 
do not introduce any noticeable changes in shapes of solitons as far as 
the widths of the soliton wavepackets are much smaller than the characteristic 
length of the traps.}.
The effective 1D coupling constants
are obtained by assuming that atoms and molecules are in the ground states of the 
2D harmonic trap of frequencies $\omega_{m,\perp}$ and $\omega_{a,\perp}$, and by
integrating the energy density over transverse variables. 
In the harmonic oscillator units, 
\bea
E_0&=&\hbar\omega_{a,x}, \cr
x_0&=&\sqrt{\frac{\hbar}{m \omega_{a,x}}}, \cr
\tau_0&=&\frac{1}{\omega_{a,x}},
\label{units}
\eea
where $E_0$, $x_0$ and $\tau_0$ are energy, length and time units, respectively,
one obtains 

\be
\lambda_a=\lambda_m=\lambda_{am} \approx 0.505,
\ee 
and 
\be
\alpha \approx 41.0.
\ee

The equations of motion for the atomic and molecular mean fields, 
corresponding to the Hamiltonian (\ref{HFR}), in the 1D model, 
in the units (\ref{units}), are
\bea
	i \frac{\partial \phi_a}{\partial t}  &=& \Big[ -\frac{1}{2} \frac{\partial^2}{\partial x^2}  
				+ \frac12 x^2 + \lambda_a N|\phi_a|^2 + \lambda_{am} N|\phi_m|^2 \Big] \phi_a \nonumber\\
				&+& \alpha\sqrt{2N}\phi_m \phi^{*}_a \cr
	i \frac{\partial \phi_m}{\partial t} &=& \Big[ -\frac{1}{4}\frac{\partial^2}{\partial x^2}  
				+ x^2 + \varepsilon + \lambda_m N|\phi_m|^2 + \lambda_{am} N|\phi_a|^2 \Big] \phi_m \nonumber\\
				&+& \alpha\sqrt{\frac{N}{2}} \phi^{2}_a,
	\label{ev}
\eea
where $N$ is the total number of atoms in the system \cite{timm}. 
The 1D detuning $\varepsilon$ is modified with respect to the 
corresponding $\cal E$ in the 3D case, i.e.
$\varepsilon = {\cal E} + (\omega_{a,\perp} - 2 \omega_{m,\perp})/\omega_{a,x}
= {\cal E}-150$.
The wave-functions $\phi_a(x)$ and $\phi_m(x)$ are normalized so that
\be
\int {\rm d}x\left(|\phi_a(x)|^2+2|\phi_m(x)|^2\right)=1.
\label{norm}
\ee
The Eqs.~(\ref{ev}) look like a pair of coupled Gross-Pitaevskii equations
but with an extra term responsible
for the transfer of atoms into molecules and molecules into atoms. The presence of this term 
allows for solutions that reveal parametric soliton like behavior. 

The mean field model (\ref{ev}) can be used if the condensates are nearly 
perfect. That is, if the depletion effects are negligible which is usually true 
even for small particle number in a system \cite{dziarmaga}. The model
neglects also effects of particle losses. The latter can be described by 
introducing terms proportional to densities with imaginary coefficients 
(that can be estimated provided there is precise experimental analysis of 
the losses in the vicinity of the Feshbach resonance)
in the square brackets of Eqs.~(\ref{ev}) \cite{yurovsky03, yurovsky04}.

\begin{figure}
\centering
\includegraphics*[width=8.6cm]{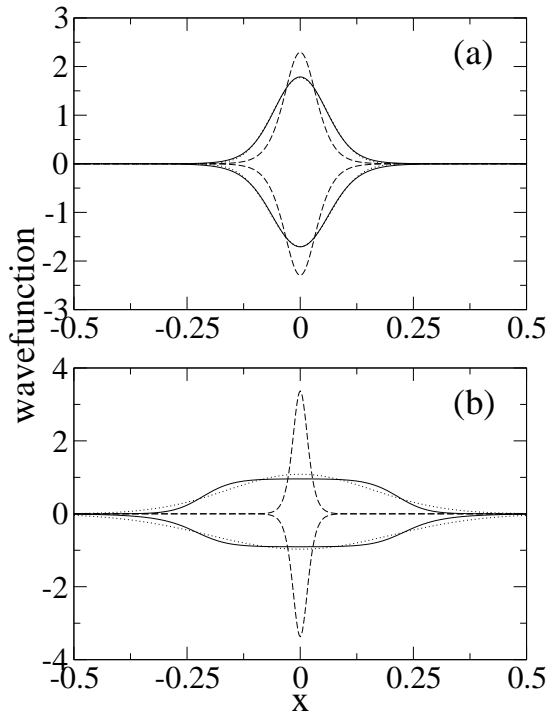}
\caption{
Solid lines correspond to the ground state solutions of Eqs.~(\ref{sev}) for $N=100$ (a) 
and $1000$ (b). Dashed lines are the analytic solutions (\ref{sol2}) corresponding to 
the case without the interaction between particles and in the absence of the trapping potentials.
Dotted lines are related to the gaussian ansatz (\ref{gau}) -- for $N=100$ the states
are almost the same as the exact ones.
The atomic wavefunctions $\phi_a > 0$, whereas the molecular ones $\phi_m < 0$. 
The ground states of Eqs.~(\ref{sev}) correspond to the detuning chosen so that 
$\langle\phi_a|\phi_a\rangle=\langle\phi_m|\phi_m\rangle$, i.e. 
$\varepsilon = -900.53$ (a) and $\varepsilon = -1505.25$ (b).
Note that the widths of the states are much smaller than the width 
of the ground state of the harmonic oscillator, which indicates that even 
for a very small particle number the nonlinearities in Eqs.~(\ref{sev}) 
determine the shapes of the states.
}
\label{ground}
\end{figure}

\section{Soliton like solutions}

\subsection{Bright soliton solutions}

The time-independent version of Eqs.~(\ref{ev}) reads
\bea
	\mu\phi_a  &=& \Big[ -\frac{1}{2} \frac{\partial^2}{\partial x^2}  
				+ \frac12 x^2 + \lambda_a N\phi_a^2 + \lambda_{am} N\phi_m^2 \Big] \phi_a \nonumber\\
				&+& \alpha\sqrt{2N}\phi_m \phi_a \cr
	2\mu\phi_m&=& \Big[ -\frac{1}{4}\frac{\partial^2}{\partial x^2}  
				+ x^2 + \varepsilon + \lambda_m N\phi_m^2 + \lambda_{am} N\phi_a^2 \Big] \phi_m \nonumber\\
				&+& \alpha\sqrt{\frac{N}{2}} \phi^{2}_a,
	\label{sev}
\eea
where $\mu$ is the chemical potential of the system, and we have assumed that 
the solutions are chosen to be real. The ground state of the system described 
by Eqs.~(\ref{sev}) can be found numerically. However, in the absence of the 
trapping potentials and for $\lambda_a = \lambda_m = \lambda_{am} =0$,
there exists an analytical solution \cite{karpierz95,drummond96,berge} if
\be
\varepsilon = -\frac{6}{l^2}, 
\ee
where
\be
l = \Big( \frac{18}{\alpha^2 N} \Big)^{1/3}.
\ee
That is,
\bea
	\phi_a(x) &=& \pm\frac {a }
	 	{ \cosh^2 \left( {\frac {x}{l}} \right)}, \cr
	\phi_m(x) &=& -\frac {a }
	 	{ \cosh^2 \left( {\frac {x}{l}} \right)},
	\label{sol2}
\eea
where 
\bea
a &=& \frac{3}{\sqrt{2 N}\alpha l^2}, \cr 
\mu&=&-\frac{2}{l^2}, 
\eea
describe the parametric bright soliton of width $l$, which 
can propagate in time without changing its shape.
The corresponding time-dependent solution is
\bea
 	\phi_a(x,t) &=& \pm\frac{a{e^{ivx}}{e^{-i( {v}^{2}+\mu) t}}} 
			{ \cosh^2 \left( {\frac {x-vt}{l}} \right)}\cr
	\phi_m(x,t) &=& - \frac{a{e^{2ivx}}{e^{- 2 i ( {v}^{2}+\mu) t}}} 
			{ \cosh^2 \left( {\frac {x-vt}{l}} \right)},
\eea
where $v$ is the propagation velocity. 
The analytical solution is known for a particular value of the detuning 
$\varepsilon$ but solutions revealing soliton like character exist also 
for other values. By changing $\varepsilon$ we can obtain solitons
but with unequal population of atomic and molecular condensates.

\begin{figure}
\centering
\includegraphics*[width=8.6cm]{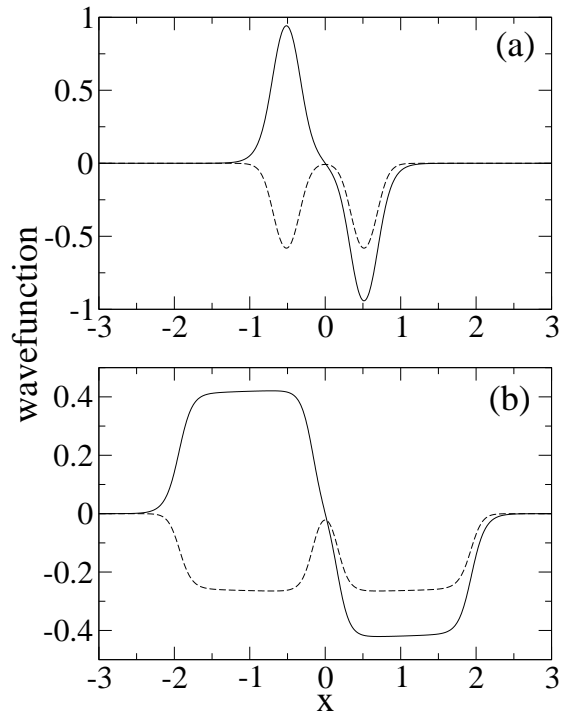}
\caption{
Two bright soliton solution for $N=100$ (a) and $N=1000$ (b). The atomic 
wave-functions $\phi_a(x)$ are odd functions while the molecular 
wave-functions $\phi_m(x)$ are the even ones. Note that populations 
of atomic and molecular condensates are unequal for the value of the detuning 
chosen, i.e. $\varepsilon = 0$.
Note that the widths of the states in panel (a) are much smaller than 
the width of the first excited state of the harmonic oscillator, which 
indicates that even for a very small particle number the nonlinearities 
in Eqs.~(\ref{sev}) determine the shapes of the states.
}
\label{2sol}
\end{figure}

In the presence of the (repulsive) interactions, the solutions for the ground 
states reveal wave-packets 
that, for increasing total number of atoms $N$ in the system, become wider 
and wider.
Indeed, for $\lambda_a>0$, $\lambda_m>0$ and 
$\lambda_{am}>0$ and with increasing $N$, the repulsive interactions are 
stronger
(see Fig.~\ref{ground}). 
Unlike the repulsive nature of the interactions, the 
term responsible for the transfer between atoms and molecules 
acts as an attractive potential. For $N$ not too high, the transfer term 
dominates over the interaction terms and it alone determines the shape of 
the wave-packets. In Fig.~\ref{ground} we can see that the ground state 
solution for $N=100$ is only slightly wider than the analytical solution 
(\ref{sol2}). 
In Fig.~\ref{ground} we also present the results of variational analysis 
of Ref.~\cite{drummondPRA04} (including also nonzero coupling constants 
$\lambda_m$ and $\lambda_{am}$) where the solutions have been assumed to be
given by gaussian functions,
\bea
\phi_a(x)=\tilde Ae^{-\tilde ax^2}, && \phi_b(x)=\tilde Be^{-\tilde bx^2}.
\label{gau}
\eea
Requiring the same fraction of atoms and molecules, for $N=100$ ($N=1000$),
we obtain $\tilde A=1.77$, $\tilde a=139.54$, $\tilde B=1.70$, $\tilde b=117.43$, 
$\varepsilon=-900$ 
($\tilde A=1.08$, $\tilde a=19.58$, $\tilde B=0.97$, $\tilde b=12.47$, 
$\varepsilon=-1429$)
which fits quite well to the exact solutions.

Even in the presence of the interactions the obtained solutions do not loose 
their solitonic character, which can be verified by the time evolution of the states
in 1D free space (i.e. in the presence of the transverse confinement but with
the axial trap turned off) --- see Sec.~IV for examples of the time evolution.
The existence of the solitons can be also verified employing the Gaussian ansatz
(\ref{gau}). Indeed, in the absence of the axial traps we have checked
that, for $N=1000$, $\lambda_m\in(0,2\lambda_a)$ and 
$\lambda_{am}\in(-2\lambda_a,2\lambda_a)$, 
the solutions exist with $\tilde a$ and $\tilde b$ in the range 
(1,20000) \cite{lam}. It shows also that the choice of equal coupling constants 
is not essential in order to deal with soliton like solutions. 

All stationary states of the Gross-Pitaevskii equations (\ref{sev}) have been
obtained numerically by means of the so-called imaginary time evolution which 
implies they are stable solutions of these non-linear coupled Schr\"odinger 
equations.

\subsection{Two bright soliton and dark soliton solutions}

In the absence of the trapping potential and for 
$\lambda_a = \lambda_m = \lambda_{am} = 0$
one can find an asymptotic solution of Eqs.~(\ref{sev}) that reveals
two bright soliton structure, 
\bea
	\phi_a(x) &=& \frac {a }
	 	{ \cosh^2 \left( {\frac {x-q}{l}} \right)} \pm \frac {a }
	 	{ \cosh^2 \left( {\frac {x+q}{l}} \right)} 
		 \cr
	\phi_m(x) &=& -\frac {a }
	 	{ \cosh^2 \left( {\frac {x-q}{l}} \right)}-\frac {a }
	 	{ \cosh^2 \left( {\frac {x+q}{l}} \right)},
\eea
valid for $q\gg l$, where
\bea
a &=& \frac{3}{\sqrt{2 N}\alpha l^2}, \cr 
\mu&=&-\frac{2}{l^2}, \cr
l &=& \Big( \frac{36}{\alpha^2 N} \Big)^{1/3}\cr
\varepsilon &=& -\frac{6}{l^2}.
\eea

In the presence of the traps and the interactions 
we may look for similar solutions. Analysis of Eqs.~(\ref{sev})
shows that due to the fact that the trapping potentials are
even functions we may expect that eigenstate functions are either even or odd.
However, because of the presence of the $\phi_a^2$ 
term in the second of Eqs.~(\ref{sev}), an even function is the only 
possibility for $\phi_m$. In Fig.~\ref{2sol} there are examples of 
the two bright soliton solutions for $N=100$ and $N=1000$. 

\begin{figure}
\centering
\includegraphics*[width=8.6cm]{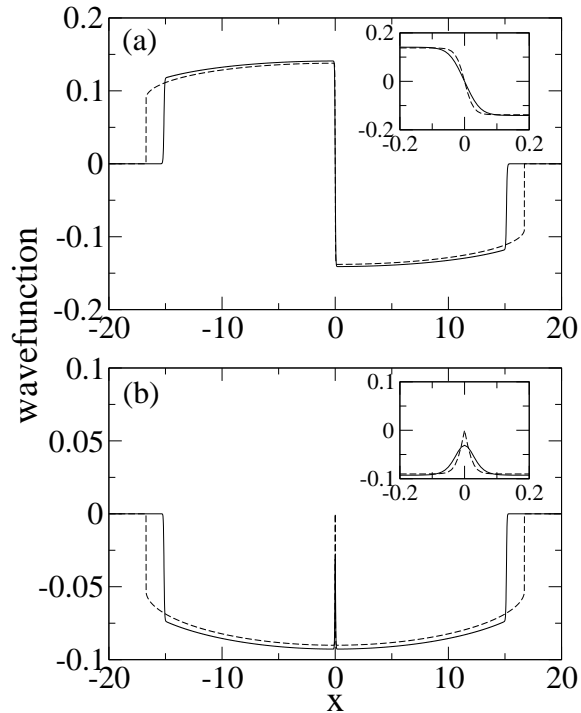}
\caption{
The soliton solution (solid lines), obtained for the detuning 
$\varepsilon=0$, compared to the solution within the Thomas-Fermi approximation
(dashed lines). Panel (a) shows the atomic wavefunctions while panel (b) the molecular ones.
The insets show the same as the main panels but with plots enlarged around the trap center.
}
\label{darksol}
\end{figure}

With an increasing number of particles the shapes of the density profiles 
change. For large particle numbers the healing length of the system 
becomes much smaller than the size of the system and one obtains density shapes 
that resemble 
the dark soliton state known for the Gross-Pitaevskii equation in the case of 
repulsive interparticle interactions \cite{burg,phil}
--- the results for $N=10^{5}$ are shown 
in Fig.~\ref{darksol}. Here, however, only  the atomic wave-function 
$\phi_a(x)$ exhibits a phase flip. We can approximate the exact eigenstates 
by (see Fig.~\ref{darksol})
\bea
	\phi_a(x)&\approx&\phi^{\rm TF}_a(x)\tanh\left(\frac{x}{\xi}\right), \cr 
	\phi_m(x)&\approx&\phi^{\rm TF}_m(x)\left|\tanh\left(\frac{x}{\xi}\right)\right|, 
\eea
where
\be
	\xi = \frac{1}{\sqrt{ \lambda_{a}N\left(|\phi_a|^2 + |\phi_m|^2 \right)}},
\ee 
and
\bea
\phi^{\rm TF}_m(x)&=& 
	\frac{\gamma}{2} - \sqrt{ \frac{\gamma^2}{4} +\frac{2\mu-x^2}{6 \lambda_{a} N} }, \cr
\phi^{\rm TF}_a(x)&=& 
	\sqrt{ \frac{2\mu-x^2}{2\lambda_{a} N} - \frac{\sqrt{2}\alpha}{\lambda_{a} \sqrt{N}} \phi^{\rm TF}_m(x)  
	-\left[ \phi^{\rm TF}_m(x) \right]^2 
	}, \cr
\gamma &=& \frac{\sqrt{2}}{3\alpha \sqrt{N}} \left(\frac{x^2-2\mu}{2} - \frac{\alpha^2}{\lambda_{a}} \right),
\eea
are solutions for the ground state of the mean field equations within the 
Thomas-Fermi approximation, and the chemical potential $\mu$ can be found from 
the normalization condition (\ref{norm}).

Finally we would like to mention that in the case of the harmonic trap, 
having any 
solutions $\phi_{a0}(x)$ and $\phi_{m0}(x)$ of the time-independent problem, 
Eqs.~(\ref{sev}), corresponding to the chemical potential $\mu$, one can easily 
obtain time evolution of the initial wave-functions $\phi_{a0}(x-q)$ and 
$\phi_{m0}(x-q)$, i.e.
\bea
\phi_{a}(x,t)&=&\phi_{a0}(x-q)e^{-i\mu t}e^{i[\dot q x-S(q)]}, \cr
\phi_{m}(x,t)&=&\phi_{m0}(x-q)e^{-i2\mu t}e^{2i[\dot q x-S(q)]},
	\label{gal}
\eea
where 
\be
S(q)=\frac12 \int_{t_0}^t {\rm d}t' \left[\dot q^2(t')-q^2(t')\right], 
\ee
and
\be
\frac{{\rm d}^2q}{{\rm d}t^2}+q=0.
\ee
The proof can be done by direct substitution of (\ref{gal}) into (\ref{ev}).
This indicates that, similarly as in the case of the Gross-Pitaevskii equation for a single condensate 
in a harmonic trap, time evolution of the translated stationary solutions reveals harmonic oscillations 
of the center of mass of the particle cloud.

\section{Experimental creation and detection}

To create experimentally a bright soliton state we propose to start with 
a purely atomic condensate in a quasi-1D trap 
with the magnetic field significantly higher 
than the resonance value. Then the field should be slowly decreased until
it reaches the resonance. Actually, from the experimental point of view,
it is better to end up slightly off the resonance because the rate of 
atomic losses is then much smaller than the rate at the resonance. 
The final populations of atomic and molecular condensates are unequal in such case 
but they still reveal solitonic character. In the following 
we analyze different velocities of the sweeping of the field in order 
to find the optimal shape of the state. 

\begin{figure}
\centering
\includegraphics*[width=8.0cm]{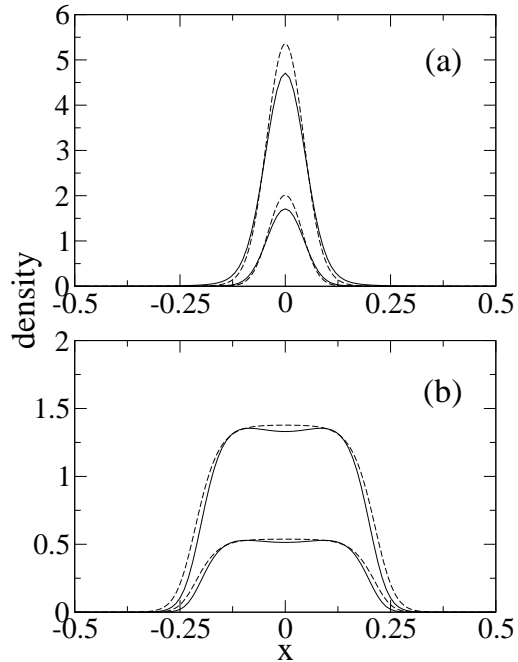}
\caption{Solid lines: density profiles of solitons excited after the magnetic field sweeping for $N=100$ (a)
and $N=1000$ (b). The sweeping lasted $95.5$~ms (a) and $93.8$~ms (b).
Dashed lines show the exact ground states corresponding to the magnetic field at the end of the sweeping. The
detuning $\varepsilon = 0$ at the end of the evolution.
Top curves, in each panel, are related to densities of atoms and bottom ones to 
densities of molecules.}
\label{evol}
\end{figure}

We have done numerical calculations for $N$ in the range between 100 and 10000. Assuming 
that the initial state is an almost purely atomic BEC, i.e. the ground state of the system far from the
resonance, the detuning $\varepsilon$ has been varied linearly in time from 3331 to 0 
($\Delta\varepsilon=3331$ corresponds to $\Delta B=0.017$~G). 
In order to obtain a nice solitonic state, the evolution time has to be carefully chosen.
We have found that the square overlap between a final state and a desired eigenstate 
oscillates as a function of the evolution time.
Therefore one has to find a proper value of the evolution time, corresponding
to a maximum value of the square overlap. From the experimental point of view, the shorter 
the evolution time the better because that allows one to reduce particle losses 
significant close to the resonance. In Fig.~\ref{evol} there are results for $N=100$ and
$N=1000$ corresponding to the optimized evolution time and compared with the  
ground states at the resonance. One can see that despite quite short evolution times
($95.5$ ms and $93.9$ ms for $N=100$ and $N=1000$, respectively), 
the final states reproduce the exact ground states quite well, i.e.
the square overlaps are $0.94$ and $0.92$. For $N=10000$ the shortest evolution 
time that results in a reasonably high squared overlap 
(i.e. $0.83$) is $90.7$~ms. 
Note that the population of the molecular BEC is smaller than the 
population of the atomic BEC because the final value of the detuning 
corresponds to the magnetic field slightly off the resonance. 

\begin{figure}
\centering
\includegraphics*[width=8.6cm]{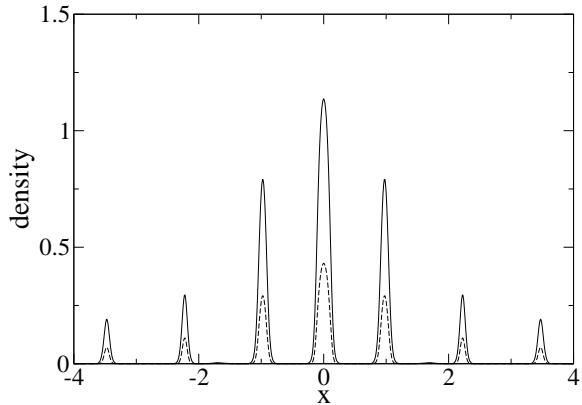}
\caption{Soliton train obtained for $N=1000$ after the sweeping of the magnetic field 
that lasted $25.5$ ms. Solid line indicates the atomic density while the dashed line 
the molecular density.
}
\label{train1000}
\end{figure}

Using the procedure described above it is possible not only to
create a single bright soliton but also excite states that reveal 
multi-peak structure 
(so-called ``soliton trains'' \cite{trains_li}). 
Actually, deviations from the optimal sweeping time of the magnetic field 
result in multi-peak structure of the atomic and molecular densities. 
An example of $N=1000$, where the soliton train is particularly well reproduced,
is shown in the Fig.~\ref{train1000}. The calculations have been performed starting with the same 
initial state and the magnetic field value as in Fig.~\ref{evol}, but the sweeping time is now
$25.5$ ms.

\begin{figure}
\centering
\begin{tabular}{c}
\includegraphics*[width=8.0cm]{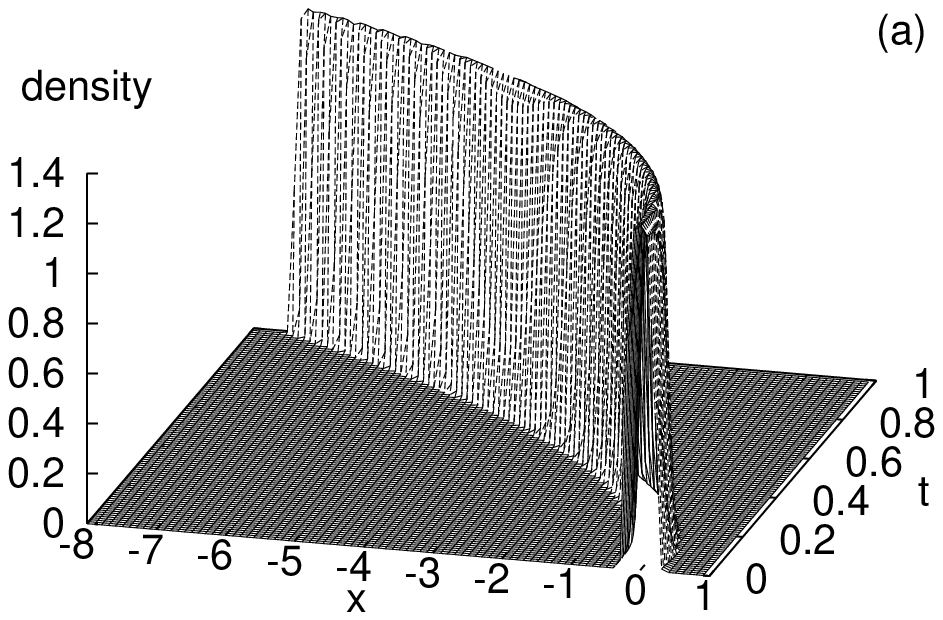}\\
\includegraphics*[width=8.0cm]{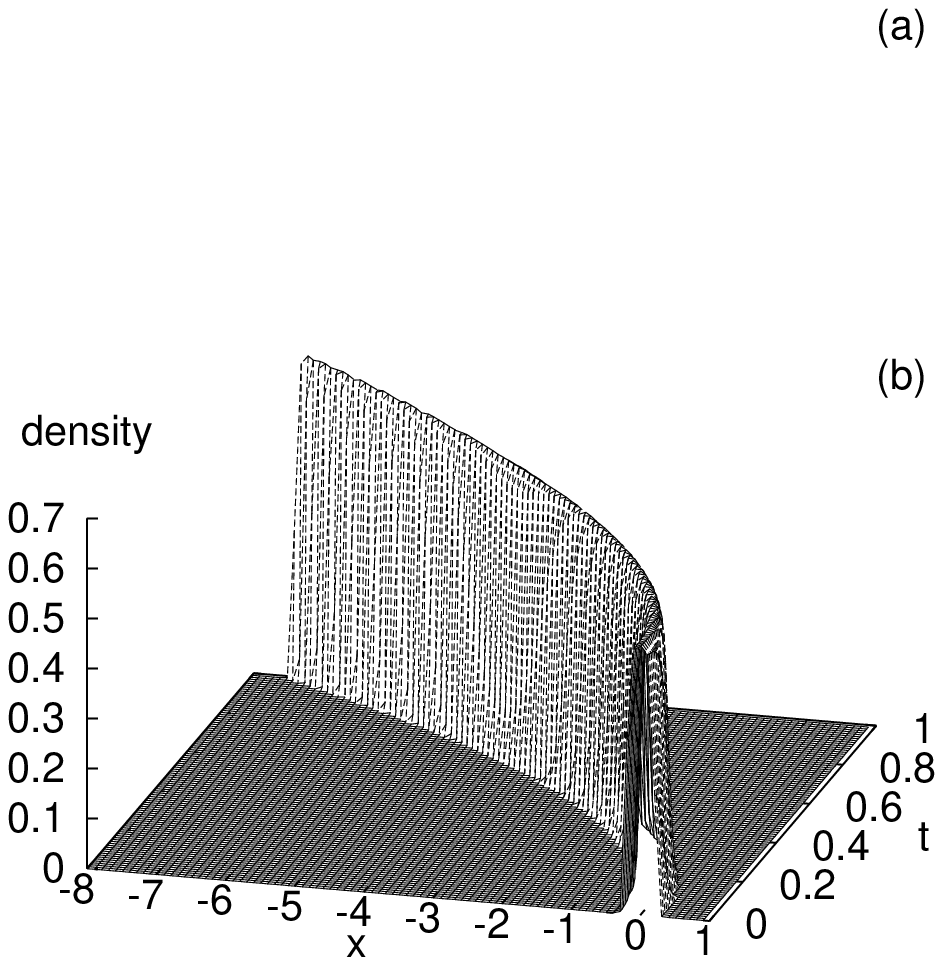}
\end{tabular}
\caption{Time evolution of a soliton state for $N=1000$ in an inhomogeneous magnetic field
        ($B_{\rm grad}=1$~G/cm), for the detuning (\ref{dtng_grad}) $\varepsilon_{\rm grad} = 0$.
	Panel (a) shows the atomic wave-packet, panel (b) the molecular one.
	The time $t=1$ in the units (\ref{units}) corresponds to $15.9$~ms.
	}
\label{gradtr}
\end{figure}

\begin{figure}
\centering
\begin{tabular}{c}
\includegraphics*[width=8.0cm]{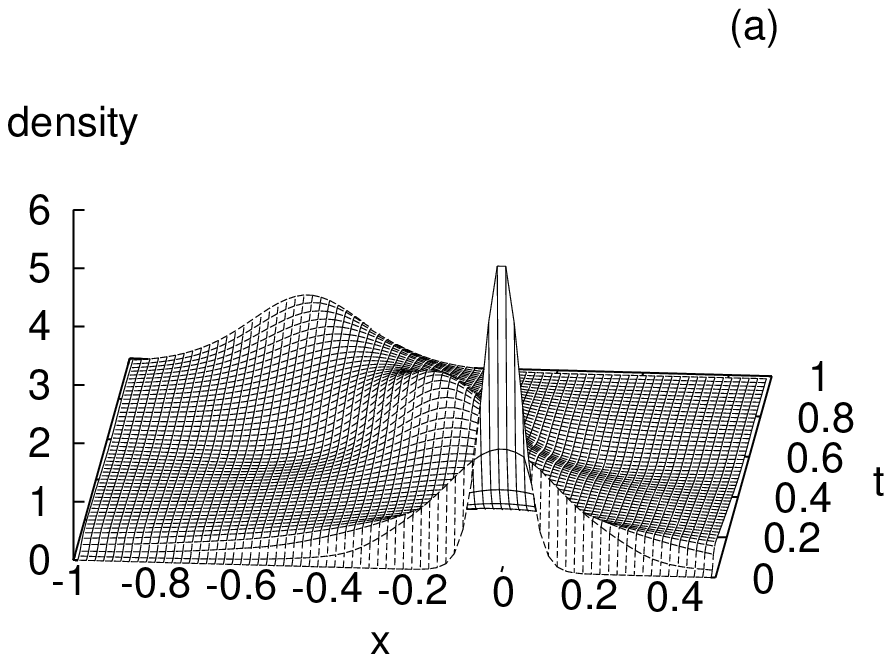}\\
\includegraphics*[width=8.0cm]{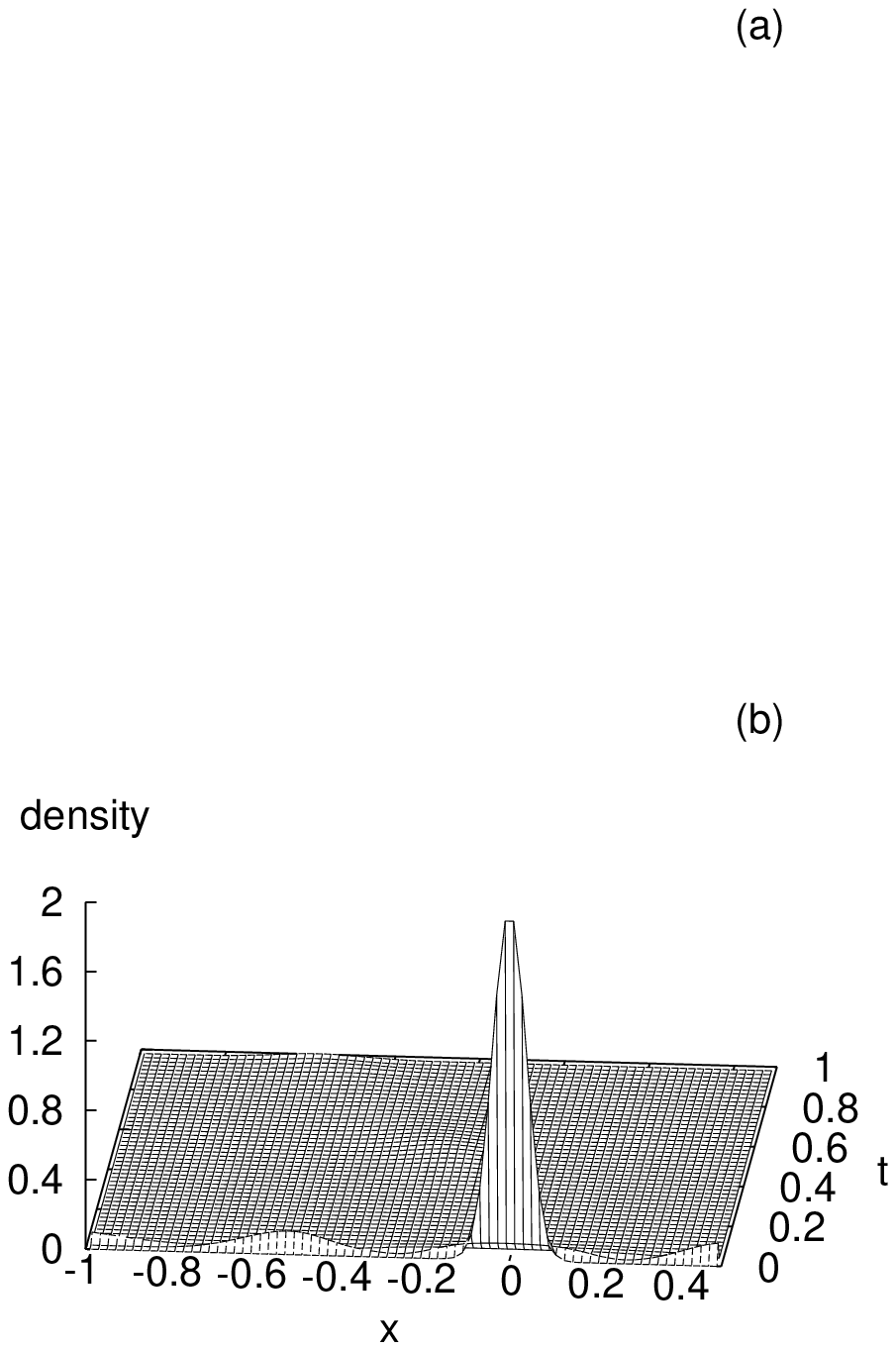}
\end{tabular}
\caption{Time evolution of a soliton state for $N=100$ in an inhomogeneous magnetic
	field ($B_{\rm grad}=1$~G/cm), for the detuning (\ref{dtng_grad}) 
	$\varepsilon_{\rm grad} = 2216$. Panel (a) shows the atomic
	wave-packet, panel (b) the molecular one. The time $t=1$ in the 
	units (\ref{units}) corresponds to $15.9$~ms.
	}
\label{gradN}
\end{figure}

\begin{figure}
\centering
\begin{tabular}{c}
\includegraphics*[width=8.0cm]{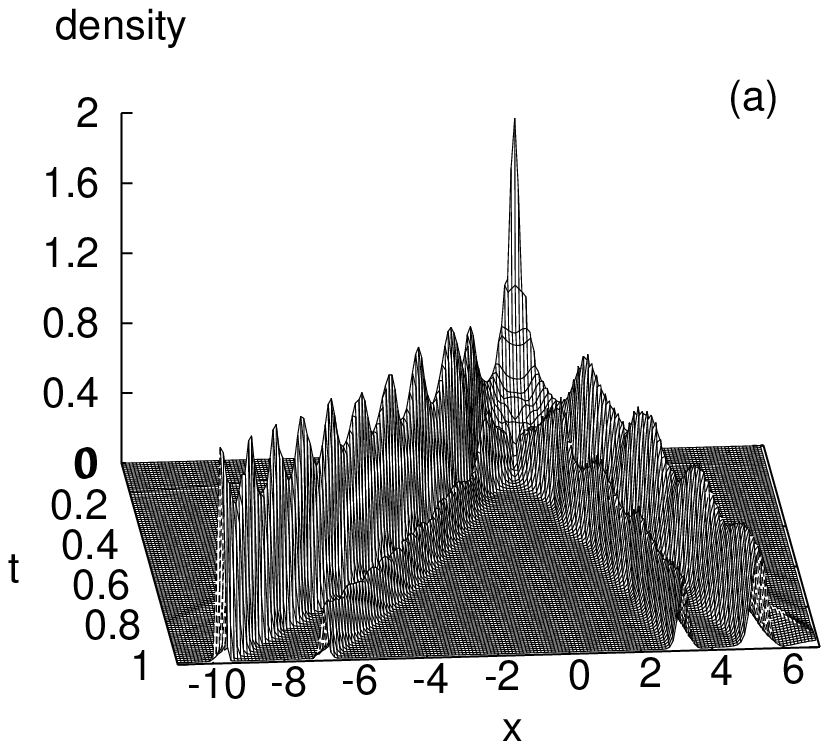}\\
\includegraphics*[width=8.0cm]{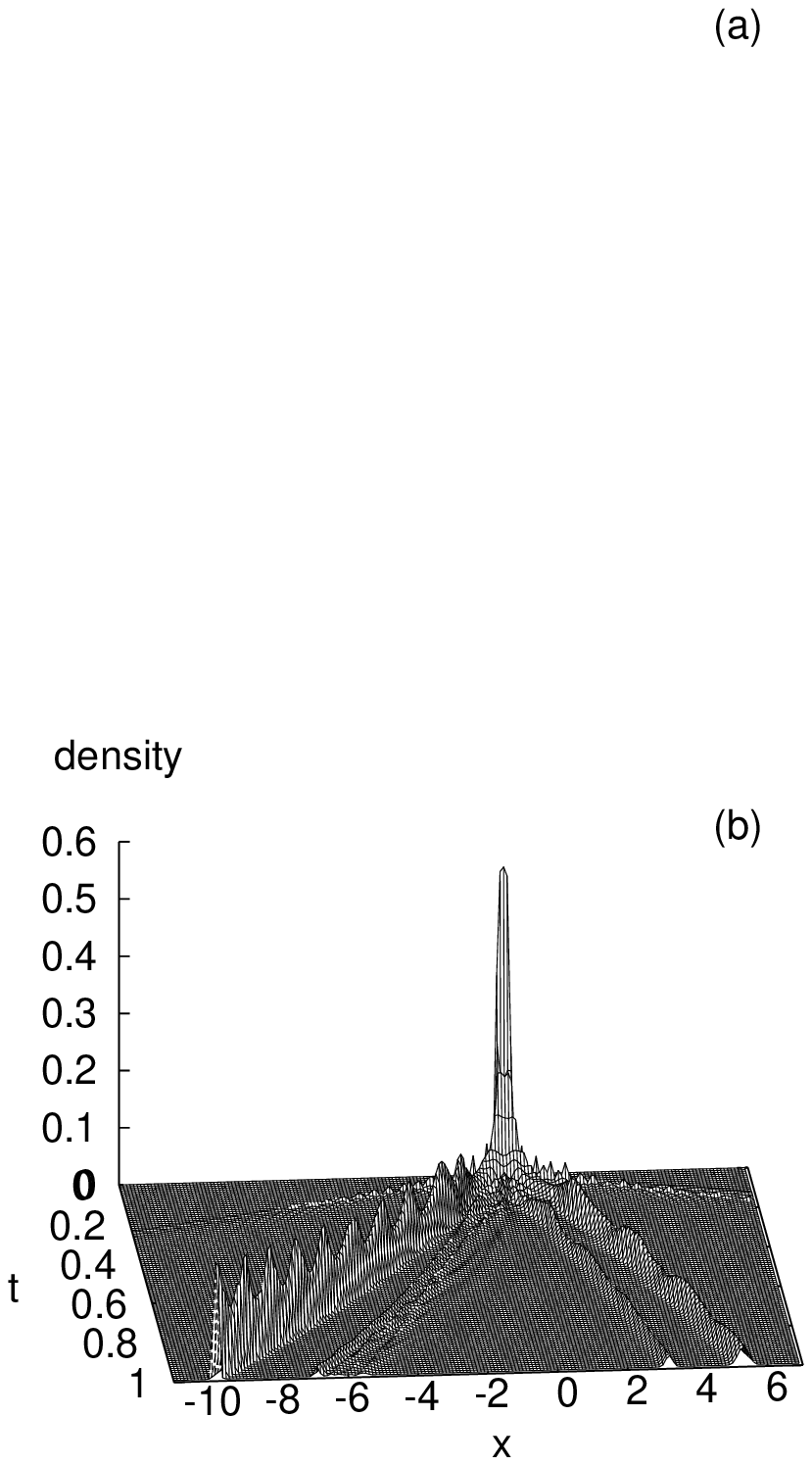}
\end{tabular}
\caption{Time evolution of a soliton state for $N=1000$ (b)
	in an inhomogeneous magnetic field ($B_{\rm grad}=1$~G/cm),
	for the detuning (\ref{dtng_grad}) $\varepsilon_{\rm grad} = 2216$.
	Panel (a) shows the atomic wave-packet, panel (b) the molecular one.
	The time $t=1$ in the units (\ref{units}) corresponds to $15.9$~ms.
	}
\label{gradN1000}
\end{figure}

In Ref.~\cite{sal,trains_li} detection of time evolution of wave-packets 
without spreading in the presence of the transverse confinement but with 
the axial trap turned off (or even in the presence of an inverted axial 
potential) was an experimental signature of the bright soliton excitation 
in an atomic BEC with attractive atom-atom interaction. 
In the present case we can use the same test to check if the states obtained 
reveal stable soliton like character. It turns out that indeed 
all the states presented do not spread when the axial trap is turned off
during time evolution lasting even $100$~ms. We do not attach the corresponding
figures because they would show practically the same density profiles as the
densities of the initial states.

In Ref.~\cite{durr_sep} the results of experimental creation of molecules 
from an atomic BEC prepared not in a quasi 1D potential but in 
a 3D trap are presented. After producing the molecules, 
the experimentalists 
analyze the evolution of the system with the traps turned off but in the presence 
of 
an inhomogeneous magnetic field. They observe separation of the molecular 
cloud from 
the atomic one, which is due to the difference between atomic and molecular 
magnetic moments.
We will analyze the evolution of the atomic and molecular clouds in an inhomogeneous
magnetic field but in the presence of the transverse confinement.
One may expect that when the solitonic states are created there is strong 
attractive coupling between atoms 
and molecules that will compete with the opposite effect resulting from 
the difference of the magnetic moments.

In the magnetic field gradient, eqs.~(\ref{ev}) are modified. Instead of the detuning
$\varepsilon$ there are 
terms proportional to the magnetic moments: $\tilde{\mu}_a(B_{grad}x_0 x + B)/E_0$
-- in the first equation, and $\tilde{\mu}_m(B_{grad}x_0 x + B)/E_0$ -- in the second one.
$B_{\rm grad}$ is a value of the magnetic field gradient and
$x_0$ and $E_0$ are length and energy units (\ref{units}).
By applying suitable unitary transformations, however, these
terms can be eliminated from the first equation. First, the coordinate
transformation $x \rightarrow x + \beta_a$ is applied, then the phases are
adjusted according to:
\bea
	\psi_a &\rightarrow& \psi_a e^{\frac{i}{2} (\beta^2_a - 
		\frac{\tilde{\mu}_a B}{E_0})t} \\
	\psi_m &\rightarrow& \psi_m e^{i(\beta^2_a - 
		\frac{\tilde{\mu}_a B}{E_0})t},
\eea
where $\beta_a = x_0 \tilde{\mu}_a B_{\rm grad} /E_0$.
Finally, there are only two additional terms in (\ref{ev})
(both present in the second of eqs.~(\ref{ev})):
\bea
	\beta x &=& \frac{x_0 \Delta \tilde{\mu}}{E_0}  B_{\rm grad} x,
\eea
and a detuning:
\bea	
\varepsilon_{\rm grad} =  \varepsilon
		- \left( \frac{x_0}{E_0} B_{\rm grad} \right)^2 \tilde{\mu}_a \Delta \tilde{\mu},
	\label{dtng_grad}
\eea
modified with respect to the case without gradient (i.e. with only $\varepsilon$).

The behavior of the system in the presence of an inhomogeneous magnetic field 
depends on the value
of the detuning $\varepsilon_{\rm grad}$. For $\varepsilon_{\rm grad} = 0$
the clouds do not separate but there is an efficient transfer of atoms into 
molecules.
In Fig.~\ref{gradtr} we see that for $N=1000$, $B_{\rm grad} = 1$~G/cm 
and $\varepsilon_{\rm grad} = 0$, the number of molecules is increased
after $15.9$~ms of time evolution.
For $\varepsilon_{\rm grad}=2216$ (corresponding to $\varepsilon=0$) 
and for small particle number (e.g. $N=100$) there is an opposite transfer,
that is of molecules into atoms (see Fig.~\ref{gradN}), and when molecules are 
absent the atomic wavepacket starts spreading. However, if $N$ is greater 
($N=1000$ or more) the wavepackets begin to split into several separated 
peaks
(see Fig.~\ref{gradN1000}) but there is still no separation between atomic and 
molecular clouds. Similar splitting of solitonic wavepackets has been analyzed 
in the case of a single component BEC in the presence of a gravitational field
\cite{theocharis}.
For greater field gradient than the value chosen in 
Figs.~\ref{gradtr}-\ref{gradN1000} or slightly different values of the coupling
constants we observe qualitatively similar behaviour.

The longer creation and detection of the solitons lasts, 
the more serious becomes the problem of atomic losses in the vicinity
of the Feshbach resonance. Measurements of the losses for different Feshbach resonances 
in $^{87}$Rb were done in a 3D trap (frequencies: $2\pi\times 50$~Hz, 
$2\pi\times 120$~Hz, $2\pi\times 170$~Hz) with an initial total number of atoms
$N = 4 \times 10^{6}$ (that corresponds to the density at the center of the atomic condensate
of order of $10^{15}$~cm$^{-3}$) \cite{durrprec}. The observed loss rate, defined as a fraction 
of atoms lost during $50$ ms hold time in the trap, was $78 \%
$ for the resonance at $B_{\rm r} = 685.43$~G. 
In the present publication we have chosen very high frequency of the transverse traps 
(i.e. $2\pi\times 1500$~Hz) in order to ensure that 
the 1D approximation is valid for all values of particle numbers considered. It 
results in the peak density for solitonic states comparable to the experimental density in \cite{durrprec}.
However, the density can be significantly reduced if one chooses weaker transverse confinement
and a moderate particle number. For example, for $N=1000$ and for the transverse traps
of $2\pi\times 200$~Hz, the peak soliton density is $5\times 10^{14}$~cm$^{-3}$ 
and the evolution time needed to obtain the final state with $0.91$ squared overlap 
with the desired eigenstate is $74.8$~ms. The losses will be smaller 
if the magnetic field at the end of the sweeping corresponds to a value 
slightly above the resonance, 
where the solitons still form but the loss rate is smaller. 

Actually, solitons can be prepared and detected in the presence of the losses
provided that the escaping atoms do not significantly excite the system of 
the remaining particles. Indeed, an experimental signature of the soliton 
excitation could be the time evolution of the system with the axial trap turned off,
where the wavepackets are expected not to increase their widths. Importantly, 
when $N$ decreases, the influence of the atom-molecule transfer term (which is 
responsible for the existence of the solitonic solutions) on the character of the
solutions grows, as compared to the influence of the interaction terms.
This is because the former depends on the particle number as $\sqrt{N}$ while the latter as $N$, 
see Eq.~(\ref{ev}). 
Consequently the normalized
particle density should not increase its width even in the presence of the
losses if a soliton state is prepared in an experiment.

\section{Summary}

We consider solitonic behaviour of coupled atomic-molecular Bose-Einstein 
condensates trapped in a quasi-1D potential. In the vicinity of a Feshbach 
resonance there are eigenstates of the system that reveal soliton-like 
character. 
The profiles of the solitons depend on the total  
number $N$ of the particles in the system and on the values of the scattering
lengths characterizing particle interactions. 
For moderate particle numbers and positive scattering lengths one 
can find two bright soliton solutions which for large $N$ turn into a state 
that resembles a dark soliton profile known for atomic condensates with the
repulsive atom-atom interaction \cite{burg,phil}. 

We analyze methods for experimental preparation and detection of the 
solitons in the atomic-molecular BECs. They can be obtained experimentally 
starting from a purely atomic BEC and slowly decreasing the magnetic field down 
to a value slightly above the resonance. Evolution times necessary to create 
solitons are quite short which is very promising experimentally because it 
allows reducing atomic losses important in the vicinity of the resonance. 
To detect the solitons one can 
perform (similarly as in Refs.~\cite{sal,trains_li}) the evolution of 
the system in the absence of the axial trap (but still in the presence of 
the transverse confinement) where the wave-packets are expected to propagate 
without spreading. 

We have also considered creation of the bright soliton trains in a coupled 
atomic-molecular 
condensate. Soliton trains in Bose-Fermi mixtures obtained by
temporal variations of the interactions between the two components has been 
recently 
proposed \cite{trainsBF}. Our system is different but its qualitative behaviour 
is to some extent similar.
In both cases, there is a term in the Hamiltonian that leads to an effective 
attractive 
interactions. This term can dominate over repulsive interactions and 
stabilize bright solitons.

\section*{ Acknowledgements }  
We are grateful to Jacek Dziarmaga and Kuba Zakrzewski for helpful discussions. 
The work of BO was supported by Polish Government scientific funds 
(2005-2008) as a research project. KS was supported by the KBN grant 
PBZ-MIN-008/P03/2003. Supports by Aleksander von Humboldt Foundation 
and within Marie Curie ToK project COCOS (MTKD-CT-2004-517186)
are also gratefully acknowledged. 
Part of numerical calculations was done in the Interdisciplinary 
Centre for Mathematical and Computational Modeling Warsaw (ICM).


\end{document}